\documentclass[a4paper,11pt]{article}
\usepackage{jheppub}
\usepackage{amsmath}
\usepackage{amssymb}
\usepackage{amsfonts}
\usepackage{slashed}
\usepackage{physics}
\usepackage{graphicx}% Include figure files
\usepackage{dcolumn}% Align table columns on decimal point
\usepackage{bm}% bold math
\usepackage{xcolor}
\usepackage{MnSymbol}
\usepackage{wasysym}

\usepackage[sort&compress]{natbib}
\title{ Realization and Stability of Non-Abelian Chiral Quantum Spin Liquids via Dimensional Reduction}

\author[a]{Rodrigo Corso}
\author[a]{Carlos A. Hernaski}
\affiliation[a]{Departamento de Física, Universidade Estadual de Londrina\\Caixa Postal 10011, 86057-970, Londrina, PR, Brasil.}

\emailAdd{rodrigocorso@uel.br}
\emailAdd{carloshernaski@utfpr.edu.br}

\abstract{	This work is concerned with the realization and stability of a non-Abelian chiral quantum spin liquid phase. To do so, we cast the problem in a quantum wires framework, which is a dimensional deconstruction framework that allows us to study the (2+1) dimensional spin liquids phase from a series of coupled (1+1) dimensional theories. The lower dimension grants us the ability to perform a bosonization procedure, which yields two different partition functions connected by a strong-weak duality transformation. This bosonization procedure is illuminating in that it makes the fixed point structure of the model unequivocal. Then, we proceed by studying the RG flow through the $ \beta $-functions, which we use to determine the phase structure. We find that the quantum spin liquid phase is realized and stable in the deep IR limit.}

\begin{document}

\maketitle
\tableofcontents
\section{Introduction} \label{Introduction}

\subsection{Motivations}	

Non-Abelian quantum matter is among the most surprising and fascinating phases of condensed matter physics. A hallmark of such phases is the presence of anyonic excitations with non-Abelian statistics \cite{moore1991nonabelions,Simon,stern2010non}. In two spatial dimensions, this is related to higher-dimensional representations of the braid group that may occur when there is a degeneracy in the spectrum. Given a degenerate state $\psi_i$ involving a set of particles, if we exchange two particles, say 1 and 2, the state changes as $\psi_i \rightarrow U_{ij}(1,2) \psi_j$, where $U_{ij}(1,2)$ is a unitary matrix. Similarly, exchanging other two particles, say 2 and 3, the state changes as $\psi_i \rightarrow U_{ij}(2,3) \psi_j$. The statistics is non-Abelian whenever $U(1,2) U(2,3)\neq U(2,3)U(1,2)$, i.e., the sequence of exchange operations does not commute. 

In addition to their spectacular physical properties, with intrinsic interest, the intensive exploration in non-Abelian states of matter is  largely due to their potentialities in applications in quantum computing \cite{Simon}. Non-Abelian excitations are expected to be supported in certain systems like in fractional quantum Hall phase (FQH), whose most celebrated representative is the Moore-Read Pfaffian state at the filling $\nu=5/2$ \cite{moore1991nonabelions}, in $p$-wave topological superconductors \cite{Read_2000}, and also in lattice spin systems (spin liquids) \cite{Levin_2005,Kitaev_2006}. 
Enlightening review articles on non-Abelian anyons can be found in  \cite{Simon,stern2010non}.

For topological phases supporting chiral massless excitations at the boundaries, the non-Abelian nature of the phase is signaled by the presence of fractionalized gapless degrees of freedom, i.e., the boundary physics corresponds to a CFT (conformal field theory) whose central charge is a non-integer rational number \cite{Kitaev_2006,Gromov_2015,Wang_2020,Bonderson_2021}. A remarkable example of the boundary of non-Abelian states, is the CFT of a Majorana fermion, which is a CFT with central charge $c=1/2$ that appears at the edge of certain non-Abelian states. This fermionic CFT is related through bosonization to the Ising CFT \cite{Seiberg:2023cdc,Shao:2023gho}, which belongs to the class of the famous unitary minimal models \cite{francesco}, characterized by the central charges
\begin{eqnarray}
	c(k)=1-\frac{6}{(k+2)(k+3)},~~~k=1,2,\cdots.
	\label{0.1}	
\end{eqnarray} 
The Ising CFT with central charge $c=1/2$ corresponds to the first member $k=1$. 

The minimal models provide an interesting setting for the construction of a whole class of non-Abelian topological phases. The central question in this undertaking is to unveil the bulk physics able to produce edge states characterized by \eqref{0.1}. Nevertheless, if we succeed in this program, we end up with a general enough series of topological phases with remarkable properties including, for example, emergent supersymmetry at the boundary, as the second member of the series \eqref{0.1} ($k=2$ - tricritical Ising model) exhibits supersymmetry \cite{Qiu:1986if}.

A very effective way to approach this problem is through the quantum wires formalism \cite{Kane2002,Teo2014,Meng:2019ket}. A method that, in a nutshell, reconstructs the bulk from the boundary degrees of freedom. The idea is to start with a set of one-dimensional decoupled CFT's carrying the degrees of freedom necessary for producing the desired edge states, and then to engineer interactions that are able to fully gap the bulk excitations while keeping some gapless degrees of freedom at the boundaries. This strategy was pursued in \cite{Huang2016}.

A key observation is that the minimal models central charges in \eqref{0.1} can be reproduced by the coset of Lie algebras
\begin{equation}
	\frac{su(2)_k\oplus su(2)_{k'}}{su(2)_{k+k'}},
	\label{0.2}
\end{equation}
with $k'=1$. This suggests that we should start with a set of CFT's involving spinful complex fermions (electrons) organized in bundles containing a number $k$ of wires, so that each bundle realizes a $u(2k)$ symmetry. According to the Sugawara construction \cite{francesco}, this symmetry structure can be decomposed as
\begin{equation}
	u(2k) \supset u(1) \oplus su(k)_2 \oplus su(2)_k.
	\label{0.3}
\end{equation} 
If we engineer interactions that gap the sectors $u(1)$ and $ su(k)_2$, we end up with a $su(2)_k$ gapless sector. Endowing the system with an additional bundle structure containing $ k'$ wires, to which we introduce similar interactions, there will remain a $su(2)_k \oplus su(2)_{k'}$ gapless sector. Then, it is possible in principle to introduce suitable interactions able to stabilize 
a two-dimensional gapped phase, leaving behind gapless modes described by the coset structure \eqref{0.2} at the boundaries. As the Abelian charge sector $u(1)$ is completely gapped in this construction, this realizes a non-Abelian chiral quantum spin liquid (QSL) rather than a quantum Hall phase.

%%%%%%%%%%%%%%%%%%%%%%%%%%%%%%%%%%%%%%%%%%%%%%%%%%%%%%%
\subsection{Statement of the Problem and Results}

After gapping the $u(1)$, $su(k)$, and $su(k')$ sectors (we will discuss the specific interactions later in the manuscript), we have a set of decoupled bundles carrying $su(2)_k \oplus su(2)_{k'}$ gapless degrees of freedom. Then we introduce interactions which, in addition to coupling the bundles to form a two-dimensional phase, open a gap in the bulk leaving behind only the desired gapless modes at the boundaries.  

\begin{figure}
	\centering
	\includegraphics[scale=.7,angle=90]{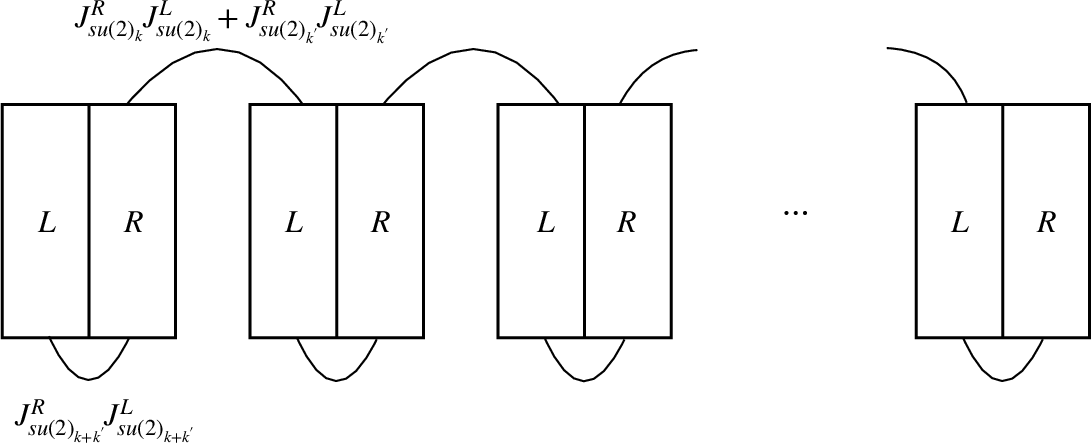}
	\caption{Schematic representation of the interactions in the quantum wires system.}
	\label{Picture}
\end{figure}

Interactions with the potential to achieve these requirements are the following. First, the current-current inter-bundle interactions, 
\begin{equation}
	\mathcal{L}_{inter} =  \sum_{i }  \lambda  \left( J^{R,i}_{su(2)_k} J^{L,i+1}_{su(2)_{k}} + J^{R,i}_{su(2)_{k'}} J^{L,i+1}_{su(2)_{k'}} \right), ~~~\lambda>0,
	\label{0.4}
\end{equation}
which are sufficient to couple the bundles and open a gap in the bulk. However, such interactions are not enough to produce the boundary modes of the form \eqref{0.2}. To this end, we add current-current intra-bundle interactions of the form
\begin{equation}
	\mathcal{L}_{intra} =  \sum_{i }  \lambda_d\, J^{R,i}_{su(2)_{k+k'}} J^{L,i}_{su(2)_{k+k'}}, ~~~\lambda_d>0,
	\label{0.5} 
\end{equation}
with the diagonal currents  
\begin{equation}
	J^{R,i}_{su(2)_{k+k'}} = J^{R,i}_{su(2)_k} +J^{R,i}_{su(2)_{k'}}~~~\text{and}~~~J^{L,i}_{su(2)_{k+k'}}= J^{L,i}_{su(2)_{k}}+J^{L,i}_{su(2)_{k'}}.
	\label{0.6}
\end{equation}
The above interactions are depicted schematically in Fig. \ref{Picture}. 

%%%%%%%%%%%%%%%%%%%%%%%%%%%%%%%%%%%%%%%%%%%%%%%%%%%%%%

Although the above interactions have the potential to produce the desired phases, there is an important issue concerning the stability of such class of non-Abelian spin liquids. This problem arises because, in general, the terms in the interactions \eqref{0.4} and \eqref{0.5} do not commute,	
\begin{equation}
	[\mathcal{L}_{intra},\mathcal{L}_{inter}]\neq 0. \label{0.7}
\end{equation}
In this way, these competing interactions may produce strong fluctuations that are able to close the gap and thus destabilize the two-dimensional phases. The expected non-Abelian phases, which require both types of interactions $\mathcal{L}_{intra}$ and $\mathcal{L}_{inter}$ operating simultaneously, may actually be unstable due to an eventual gap closing. Our main objective in this work is to investigate the stability of such phases. 

\begin{figure}
	\centering
	\includegraphics[scale=.7,angle=90]{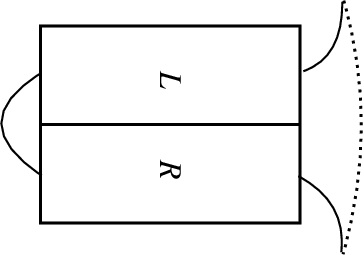}
	\caption{Single-bundle, representing a typical bundle of the bulk.}
	\label{Singlebundle}
\end{figure}

The physics we are interested in can be isolated by considering a single-bundle system with periodic boundary conditions, as shown in Fig. \ref{Singlebundle}. It represents a typical bundle of the bulk, where the two types of interactions  $\mathcal{L}_{intra}$ and $\mathcal{L}_{inter}$ are competing. This system embodies the physical mechanism \eqref{0.7}, while at the same time it considerably simplifies the whole analysis. The stability of the non-Abelian spin liquids is equivalent to the opening of a gap in the single-bundle system in the infrared (IR).

We carry out a {\it non-perturbative} renormalization group study of the single-bundle system and compute the beta functions of the competing coupling constants. To this end, we bosonize the original fermionic model and  obtain a set of coupled Wess-Zumino-Witten (WZW) theories. Then, by using the methods developed in a set of recent works \cite{Georgiou2017, Georgiou:2018vbb, Delduc:2020vxy, Sfetsos:2013wia, georgiou2017integrable, Georgiou2017a, Georgiou2020, Georgiou2015}, we are able to compute the beta functions in the large level limit. The main result of this work is the verification of the stability of non-Abelian topological phases driven by the interactions \eqref{0.4} and \eqref{0.5}, which follows from the numerical solution of the beta functions. 

This work is organized as follows. Section \ref{sec2} is dedicated to presenting the model, its symmetry content, and the proposed interactions. Section \ref{first_bosonization} focuses on two different, but equivalent, bosonization procedures, while section \ref{duality_section} is concerned about the resulting duality transformation that maps between the two resulting partition functions. In section \ref{Fixed_Points}, we discuss the fixed point structure of the model from one bosonic partition function and comment on the phase structure of the complete wire model with the bundle structure. We proceed to section \ref{RG flow}, where we study the RG flow of the model through its beta functions, discuss the realization of the fixed points and the phase structure of the model. At last, we present our final considerations in the section \ref{final remarks}.

\section{The Model} \label{sec2}

We consider a system consisting of one bundle of $N_1+N_2$ wires. For the sake of generality, we allow a set $N$ spinfull fermions to propagate in each wire. In particular, $N=2$ reproduces the minimal model structure of \eqref{0.2}.  
Therefore, the basic ingredients in the construction are complex fermions equipped with the following index structure
\begin{equation}
	\psi_{R/L, i\,  \sigma_I}^{(I)},
\end{equation}
with  $i=1,\ldots,N$, $\sigma_I=1,\ldots,N_I$, and $I=1,2$, and governed by the action
\begin{align}
	S_{0}=\int \dd[2]{z} \; \left[\tilde{\psi}_{R,i \sigma_1}^{(1)\dagger}\partial\tilde{\psi}^{(1)}_{R i \sigma_1}+\tilde{\psi}_{L,i \sigma_1}^{(1)\dagger}\bar \partial\tilde{\psi}^{(1)}_{L i \sigma_1}+\tilde{\psi}_{R,i \sigma_2}^{(2)\dagger }\partial\tilde{\psi}^{(2)}_{R i \sigma_2}+\tilde{\psi}_{L,i \sigma_2}^{(2)\dagger}\bar\partial\tilde{\psi}^{(2)}_{L i \sigma_2}\right]\label{4.1}.
\end{align}

The action is invariant under transformations of the symmetry groups $ U_{R}(N N_1)\times U_{L}(N N_1)\times U_{R}(N N_2)\times U_{L}(N N_2) $, with the associated Lie algebras
\begin{align}
	u(N N_1)&\supset  u(1)\oplus su(N)_{N_1}\oplus su(N_1)_{N}, \nonumber\\
	u(N N_2)&\supset  u(1)\oplus su(N)_{N_2}\oplus su(N_2)_{N}. \label{4.2}
\end{align}
Then, according to the Sugawara construction, the corresponding energy-momentum tensors can be decomposed as
\begin{align}
	T[u(NN_1)]&=T[u(1)]+T[su(N)_{N_{1}}]+T[su(N_{1})_{N}], \nonumber\\
	T[u(N N_2)]&=T[u(1)]+T[su(N)_{N_{2}}]+T[su(N_{2})_{N}], \label{sugawara}
\end{align}
which is reflected on the splitting of the central charges
\begin{align}
	c_{\psi^{(1)}}&=NN_{1}=1+\frac{N_{1}\left(N^{2}-1\right)}{N+N^{1}}+\frac{N\left(N_{1}^{2}-1\right)}{N+N_{1}},\nonumber\\	c_{\psi^{(2)}}&=NN_{2}=1+\frac{N_{2}\left(N^{2}-1\right)}{N+N_{2}}+\frac{N\left(N_{2 }^2-1\right)}{N+N_{2}}. \label{fermion_central_charges}
\end{align}

As discussed in the Introduction, in order to realize the $ (2+1) $-dimensional non-Abelian spin liquids, we introduce operators that selectively open a gap in each one of the sectors of $ u(N N_1) $ and $ u(N N_2)$, so that the resulting phase is fully gapped. Such interactions were constructed in \cite{Huang2016} and will be described in the following.

\subsection{U(1) sector}

We start by discussing the interactions in the $U(1)$ charge sector. First, the symmetry structure of the model allows a $U(1)$ current-current Thirring interaction of the form 
\begin{align}
	\mathcal{L}_{Thirring}=\sum_{I=1,2}\lambda_{I}^{t}J_{R}^{(I)}J_{L}^{(I)}, \label{Thirring_Interactions}
\end{align} 
where $ J^{(I)}_{R/L}=\sum_{i,\sigma_I}\psi_{R/L;i \sigma_I}^{(I)\dagger}\psi^{(I)}_{R/L;i \sigma_I}$. Even though this interaction is exactly marginal and hence does not open a gap in the charge sector, we shall consider it since it may be important to make other interactions relevant. The interaction responsible for opening a gap in the $U(1)$ sector is a generalized version of the 
Umklapp interaction, given by
\begin{align}
	\mathcal{L}_{Umklapp}&=-\sum_{I=1,2}{\lambda}_I^{u}\left(\prod_{i=1}^{N}\prod_{\sigma=1}^{N_{I}}\tilde{\psi}_{R i \sigma}^{(I)\dagger} \right) \left(\prod_{i=1}^{N}\prod_{\sigma=1}^{N_{I}}\tilde{\psi}^{(I)}_{L i \sigma}\right)+\left(R \leftrightarrow L\right). \label{Umklapp_interactions}
\end{align}
In appendix \ref{AA}, we show through Abelian bosonization that the charge degrees of freedom completely decouple from the model and this interaction is equivalent to a Sine-Gordon interaction, which opens a mass gap in IR. Therefore, as regarding the stability of the low-energy phase, we can concentrate only in the non-Abelian sectors of the model.

\subsection{Non-Abelian sectors}

Non-Abelian current-current interactions are perturbatively IR-relevant and thus may open a stable gap in the corresponding sectors \cite{Huang2016}. For the $SU(N_{I})$ sectors, we consider 
\begin{align}
	\mathcal{L}_{SU(N_{1})}+\mathcal{L}_{SU(N_{2})}=\sum_{A,I}\frac{\tilde{g}^{I}}{N} J_{R}^{(I)A}J_{L}^{(I)A}, \label{4.3}
\end{align}
where the $ SU(N_{I}) $ currents are given by $ J^{(I)A}_{R/L}=\sum_{i,\sigma,\rho}\psi_{R/L; i\sigma}^{(I)\dagger}t^{(I)A}_{\sigma \rho} \psi_{R/L; i \rho}^{(I)}  $, and $ t^{(I)A} $ are the group generators. We use the following conventions
\begin{align}
	\begin{array}{l}
		{\left[t^{A}, t^{B}\right]=i f^{ABC} t^{C}, \quad \tr\left(t^{A} t^{B}\right)=\frac{1}{2}\delta^{AB}},\quad \text{and} \quad f^{A CD} f^{B CD }=N_I \delta^{A B}
		\label{4.4.1}
	\end{array}
\end{align}
with $ A,B=1, \ldots, N_I^{2}-1$ and $I=1,2$. 

For this single-bundle case, the inter-bundle interactions in \eqref{0.4} reduce to 
\begin{align}
	\mathcal{L}_{SU(N)}=\sum_{a,I}\frac{\lambda_I}{N_I}J^{(I) a}_{R}J^{(I) a}_{L},\label{4.5}
\end{align}
where the $ SU(N) $ currents are given by $ J_{R/L}^{(I) a}=\sum_{i,j,\sigma}\psi^{(I)\dagger}_{R/L; i \sigma}t^{a}_{ij}\psi^{(I)}_{R/L; j \sigma} $, with $a=1,\ldots,N^2-1$, with the same normalization as in \eqref{4.4.1}, whereas the diagonal intra-bundle interactions \eqref{0.5} become
\begin{align}
	\mathcal{L}_{d}= \frac{\lambda_{d}}{N_1+N_{2}}\sum_a K^{a}_{R}K^{a}_{L}, 
\end{align}
with $K_{R/L}^a=J^{(1)a}_{R/L}+J^{(2)a}_{R/L}$.

To proceed with our discussion we consider the quantum partition function for this interacting model with a more compact notation
\begin{align}
	Z=\int \mathcal{D}\psi\exp - &\int \dd[2]{z}\left[\mathcal{L}_{0}+\mathcal{L}_{Thirring}+\mathcal{L}_{Umklapp}\right. \nonumber \\
	& + \left.\mathcal{L}_{SU(N_1)}+\mathcal{L}_{SU(N_2)}+8\pi\sum_{IJ}\frac{\tilde{\lambda}_{IJ}}{\sqrt{N_I N_J}}J^{(I)a}_R J^{(J)a}_L\right], \label{interacting model}
\end{align}
where
\begin{equation}
	\tilde{\lambda}_{IJ}=\left(\begin{array}{cc}
		\lambda_1 +N_1\lambda_d& \sqrt{N_1N_ 2}\lambda_d\\
		\sqrt{N_1N_ 2}\lambda_d& \lambda_2 +N_2\lambda_d
	\end{array}\right).
\end{equation}
Our main objective is to unveil the renormalization group flow of the couplings $\lambda_{1},\lambda_{2}$, and $\lambda_{d}$, to analyze the stability of the non-Abelian topological phases.

\section{Bosonization} \label{first_bosonization}

The first step to carry out the renormalization group analysis is to consider the bosonized version of the model \eqref{interacting model}. To this end, we rewrite the partition function by introducing auxiliary vector fields valued in the Lie algebras associated with each group structure of our model \eqref{4.2}. This gives
\begin{align}
	S=\int \dd[2]{z}&\left[\sum_I\left(\psi_{R}^{(I) \dagger}D ^{I}_{\bar z} \psi^{(I)}_{R}
	+\psi_{L}^{(I)\dagger}  D^{I}_{ z}\psi^{(I)}_{L}+\frac{N}{4\pi\tilde g^I}\tr C^I_{z}C^I_{\bar z}+ \frac{1}{\lambda^I_{t}}A^I_{z}A^I_{\bar z}\right)\right.\nonumber\\
	& \left. +\sum_{IJ}\frac{\sqrt{N_IN_J}}{4\pi}\tilde{\Gamma}^{IJ}\tr B^I_{z}B^J_{\bar z}
	+\mathcal{L}_{Umklapp}\right] \label{action with gauge fields},
\end{align}
where $\tilde{\Gamma}_{IJ}=\tilde{\lambda}^{-1}_{IJ}$ and the covariant derivatives are defined by
\begin{align}
	D ^{I ij;\sigma_I \rho_I}_{\mu}&=(\partial_{\mu}-iA^I_{\mu})\delta_{ij}\delta^I_{\sigma_I \rho_I}-i(B^{I a}_{\mu})t^{a}_{ij}\delta^I_{\sigma_I\rho_I}-iC^{I A}_{\mu}t^{I A}_{\sigma_I\rho_I}\delta_{ij},
\end{align}
The vector fields $ A^I_{\mu} $,  $ B^I_{\mu}=-iB^{I a}_{\mu}t^{a}_{ij} $, and $ C^I_{\mu}=-iC_{\mu}^{I A}t^{I A}_{\sigma_I\rho_I}$,  are $ U(1) $, $ SU(N) $, and $SU(N_I) $  gauge fields, respectively. The indices $\sigma_I$ and $\rho_I$ in $\delta^{I}_{\sigma_I\rho_I}$ and $t^{I A}_{\sigma_I\rho_I}$ run from one to $N_I$. By integrating out the vector fields we recover the original action \eqref{interacting model}.

In order to reproduce the free fermion as we turn off the coupling constants, we shall impose that in the free fermion limit all the vector fields go to zero. In this way, their integration and the free fermion limit do commute and we can safely discuss RG flows starting at the free fermion fixed point.

We proceed by performing a change of variables that decouples the vector gauge fields, associated with each group structure of \eqref{4.2}, from the fermions fields \cite{Naon:1984zp}. For the non-Abelian sectors, this is achieved with the field redefinitions
\begin{align}
	\psi^I_{L;i\sigma}&\rightarrow U^I_{ij}V^{I}_{\sigma\rho}\psi^I_{L;j,\rho}, \qquad \psi_{L;i,\sigma}^{I \dagger}\rightarrow \psi_{L;j,\rho}^{I\dagger}U^{I-1}_{ji}V^{I-1}_{\rho\sigma},\\
	\psi^I_{R;i\sigma}&\rightarrow U^{I\dagger -1}_{ij}V^{I\dagger -1}_{\sigma\rho} \psi^I_{R;j,\rho}, \qquad \psi_{R;i\sigma}^{I\dagger}\rightarrow \psi_{R;j,\rho}^{I\dagger}U^{I\dagger}_{ji}V^{I \dagger}_{\rho\sigma},\label{rep2}
\end{align}
where $U^I$ and $U^{I\dagger}$ are $ SU(N) $ fields,  whereas $V^I$ and $V^{I\dagger}$ are $ SU(N_I) $ fields. The decoupling emerges when we parametrize the gauge fields in terms of the group-valued fields as
\begin{eqnarray}
	C^I_{\bar z}&=&-V^{I \dagger -1}\bar \partial V^{I \dagger}, \qquad C^I_{z}=\partial V^I V^{I-1},\\
	B^I_{\bar z}&=&-U^{I \dagger -1}\bar \partial U^{I\dagger}, \qquad B^I_{z}=\partial U^I U^{I-1}.\label{rep1}
\end{eqnarray}
The Umklapp interaction remains invariant under the $SU(N)$ transformations.

Furthermore, from the non-Abelian chiral anomaly, the fermion field reparametrizations generate the Jacobians \cite{Witten1984,Fradkin1987}
\begin{align}
	J_{F}[\psi_{R/L}]&=e^{\sum_I\left[N\left(W[V^{I\dagger}]+W[V^I]+\frac{c}{\pi}\int d^2z\tr C^I_{\bar{z}}C^I_z \right)+N_I\left(W[U^{I\dagger}]+W[U^I]+\frac{b}{4\pi}\int d^2z\tr B^I_{\bar{z}}B^I_z \right)\right]}\label{fermionic jacobian},
\end{align}
with $ W[g] $ being the Wess-Zumino-Witten action
\begin{align}
	W[g]=\frac{1}{8\pi}\int_{\partial \mathcal{M}}\dd[2]{z}\tr\partial_{z}g\partial_{\bar z}g^{-1}-\frac{i}{48\pi}\int_{\mathcal{M}} \dd[3]{z}\epsilon^{\mu \nu \sigma}\tr g^{-1}\partial_{\mu}gg^{-1}\partial_{\nu}gg^{-1}\partial_{\sigma}g. \label{wzw action}
\end{align}
The parameters $b$ and $c$ incorporate the regularization ambiguities of the fermionic determinant.

Putting everything together, the full partition function associated with the action \eqref{action with gauge fields} can be cast into the form
\begin{align}
	Z=Z_{U(1)}Z_{SU(N_1)}Z_{SU(N_2)}Z_{SU(N)}\label{4.8},
\end{align}
with
\begin{align}
	Z_{U(1)}=&\int \mathcal{D} A\mathcal{D} \psi\exp -\int \dd[2]{z}\left[\sum_I\left(\psi_{R,i \sigma}^{I\dagger}D_{\bar z}(A^I)\psi^I_{R,i\sigma}+\psi_{L,i \sigma}^{I\dagger}D_{z}(A^I)\psi^I_{L,i\sigma}
	+\frac{1}{\lambda^I_{t}}A^I_{z}A^I_{\bar z}\right)\right.\nonumber\\+&\left.\mathcal{L}_{Umklapp}\right]\label{4.9},
\end{align}
\begin{align}
	Z_{SU(N_{I})}&= \int \mathcal{D} C\exp  N\sum_I\left[W[V^{I\dagger}]+W[V^I]+\frac{1}{4\pi}\left(\frac{1}{\tilde{g}^I}+c\right)\int \dd[2]{z} \tr C^I_{\bar{z}}C^I_z\right], \label{4.10}
\end{align}
and
\begin{align}
	Z_{SU(N)}&= \int \mathcal{D} B\exp \left[\sum_I N_I\left(W[U^{I\dagger}]+W[U^I]\right)+\frac{1}{4\pi}\sum_{I,J}\sqrt{N_I N_J}\Gamma^{IJ}\int \dd[2]{z} \tr B^I_{\bar{z}}B^J_z\right], \label{4.11}
\end{align}
with 
\begin{align}
	\Gamma^{IJ}=\tilde{\Gamma}^{IJ}+b\delta^{IJ} \label{gamma}.
\end{align}

The form of the total partition function \eqref{4.8} shows us that we can treat each group sector separately. The Abelian sector containing the Umklapp interaction will be discussed in detail in  Appendix \ref{AA}, where we show that this sector possesses the free fermion fixed point at UV and the interaction gives a gap for two degrees of freedom related to charged $U(1)$ sectors. The behavior of the non-competing $ SU(N_{I}) $ sectors, described by \eqref{4.10}, was recently discussed in \cite{Santos2023}, so that we will restrict ourselves to drawing comments on it. The dynamical behavior of the competing interactions involving the $SU(N)$ currents are then completely characterized in the terms of the bosonic model $Z_{SU(N)}$ in \eqref{4.11}. This model has the form of a set of WZW models with distinct arbitrary (and negative) levels perturbed by generic quartic interactions (quadratic in the currents).

%%%%%%%%%%%%%%%%%%%%%%%%%%%%%%%%%%%%%%%%%%%%%%%%%%%%%%

\section{Duality} \label{duality_section}

One important consistency check of the bosonized model \eqref{4.11} can be done via duality. The model is expected to exhibit a dual version with positive levels via a weak-strong relation \cite{Santos2023,Kutasov1989}. To show this feature, we present the model \eqref{4.11} in two distinct but equivalent forms. 

To present the first form, let us multiply and divide the partition function \eqref{4.11} by the following factor
\begin{eqnarray}
	Z^2_{N_1}Z^2_{N_2}=\prod^2_{I=1}\int Dg^I e^{-N_IW[g^{I}]}\int Dg^{I\dagger} e^{-N_IW[g^{I\dagger}]},
\end{eqnarray}
with $g^I$ and $g^{I\dagger}$ being $SU(N)$ fields. Then, we perform the change of variables $g^I\rightarrow U^{I\dagger}g^I$, $g^{I\dagger}\rightarrow g^{I\dagger}U^{I}$, and
make use of the identity
\begin{eqnarray}
	W[U^{I\dagger} g^I]+W[g^{I\dagger}U^I]-W[U^{I\dagger}]-W[U^I]&=&W[g^{I\dagger}]+W[g^I]+\\\nonumber
	&-&\frac{1}{4\pi}\int d^2z\tr\left(B^I_zg^{\dagger-1}_I\partial_{\bar{z}}g^\dagger_I-B^I_{\bar{z}}\partial_{z}g_Ig^{-1}_I\right),
\end{eqnarray} 
to obtain
\begin{eqnarray}
	Z_{SU(N)}&=&\frac{1}{Z^2_{N_1}Z^2_{N_2}}\int \mathcal{D} B^I\mathcal{D} g^I\mathcal{D}g^{I\dagger}\exp -\sum_I N_I\left[W[g^{I\dagger}]+W[g^I]\right.\\\nonumber
	&-&\left.\frac{1}{4\pi}\int d^2z\tr\left(B^I_zg^{\dagger-1}_I\partial_{\bar{z}}g^\dagger_I-B^I_{\bar{z}}\partial_{z}g_Ig^{-1}_I+\sum_{J}\sqrt{\frac{N_J}{N_I}}\Gamma^{IJ}B^I_{\bar{z}}B^J_z\right)\right].
\end{eqnarray}
Integrating over the gauge fields, we get
\begin{eqnarray}
	Z_{SU(N)}&=&\frac{1}{\left(Z_{N_1}Z_{N_2}\right)^2}\int\mathcal{D} g^I\mathcal{D}g^{I\dagger}\exp -\left[\sum_I N_I\left(W[g^{I\dagger}]+W[g^I]\right)\right. 	\label{bos1} \\\nonumber
	&+&\left.4\pi\sum_{IJ}\sqrt{N_I N_J}\lambda_{IJ}\int d^2z\tr\left(J^I_R(g)J^J_L(g^{\dagger})\right)\right],
\end{eqnarray}
where we have defined $\lambda\equiv \Gamma^{-1}$ and
\begin{equation}
	J_R(g)=\frac{1}{4\pi}\partial_{z}gg^{-1} \qquad J_L(g^{\dagger})=-\frac{1}{4\pi}g^{\dagger-1}\partial_{\bar{z}}g^\dagger.
	\label{currents}
\end{equation}
The model \eqref{bos1} can be understood as the bosonized version of the competing sector of the fermionic interactions. We note that it involves a set of coupled WZW theories with positive level and, consequently, is well-defined in the perturbative regime. As we shall see, this form is also convenient to analyze the RG flow of the coupling constants. 

We can instead make a direct change of variables from $B^{I}_{\bar{z}}, B^{I}_{z}$ to $U^{I\dagger}, U^I$ in \eqref{4.11} given by \eqref{rep1}. As is well known \cite{Polyakov:1983tt,polyakov1984goldstone,Cabra1990}, this change of variables generates a non-trivial Jacobian given by a WZW term:
\begin{align}
	J_{B}&=\det \left(D_{z}(B^I)D_{\bar z}(B^I)\right)_{adj}=\det\left(\partial_{z}\partial_{\bar z}\right)_{adj}e^{2N W\left([U^{I \dagger}]+W[U^I]\right)}, \label{bosonic jacobian} 
\end{align}
where the subscript $adj$ means that the determinant is taken in the adjoint representation and the remaining determinant can be expressed in terms of ghost fields. Possible regularization ambiguities are incorporated in the parameter $b$ in the definition of $\Gamma$ in \eqref{gamma}. Besides the non-trivial Jacobian, the change of variables also generates a further renormalization factor $f^{IJ}$ for the currents coupled to $A^I_{\bar{z}}$ and $A^I_z$. Taking this possible renormalization into account, we get
\begin{eqnarray}
	Z_{SU(N)}&=& Z_g^{2}\int \mathcal{D} U^I\mathcal{D} U^{I\dagger}\exp \left[\sum_I\left(N_I+2N\right)\left(W[U^{I\dagger}]+W[U^I]\right)\right.\nonumber\\
	&+&\left.4\pi\sum_{I,J}\sqrt{N_I N_J}f_{IJ}\Gamma^{IJ}\int \dd[2]{z} \tr J^I_L(U^\dagger)J^J_R(U)\right], \label{4.12}
\end{eqnarray}
where $Z_g=\det\left(\partial_{z}\partial_{\bar z}\right)_{adj}$ is a partition function that can be expressed in terms of an action
for ghost fields, 
\begin{eqnarray}
	S_{ghost}= \int d^2z\sum^{N^2-1}_{i=1}\left(b^i_z\partial_{\bar{z}}\bar{c}^i+b^i_{\bar{z}}\partial_zc^i\right). \label{4.121}
\end{eqnarray}
The ghost fields $b^i_{z}$ and $b^i_{\bar{z}}$ have conformal weight one, while the fields $c^i$ and $\bar{c}^i$ have conformal weight zero. The partition function in \eqref{4.12} involves a set of coupled WZW theories with negative level, and then it is not well-defined in the perturbative regime.  

Now we expect some duality between \eqref{bos1} and \eqref{4.12} since both were derived from the same partition function  \eqref{4.11}. Comparing them, the first point to notice is that they would be related though a level reflection plus a shift $N_I\rightarrow - N_I-2 N$. The fact that theory  \eqref{bos1}  is well-defined perturbatively suggests that this level changing should be accompanied by a changing in the coupling constants $\lambda \rightarrow \lambda^{-1}$, so that we end up with a weak/strong-like duality. 
Then, we take the duality transformations as 
\begin{eqnarray}
	N_I&\rightarrow& - N_I-2 N,\nonumber\\
	\lambda&\rightarrow& \lambda^{-1}. \label{4.14}
\end{eqnarray} 
In mapping \eqref{bos1} into \eqref{4.12} using these transformations, we need to be careful with the term $\sqrt{N_I N_J}$, which must be understood as 
\begin{eqnarray}
	\sqrt{N_I} \sqrt{N_J} &\rightarrow& \sqrt{-(N_I+2N)} \sqrt{-(N_J+2N)}\nonumber\\
	&\rightarrow& -  \sqrt{(N_I+2N)} \sqrt{(N_J+2N)},
\end{eqnarray}
in order to properly take into account  the sign change. 

Now we have to choose the renormalization factors $f_{IJ}$ so that upon these transformations the theory \eqref{bos1} is mapped in \eqref{4.12}. The delicate point here is that the matrix coupling constants $\lambda$ is level-dependent. To take this into account, we choose the renormalization factors $f_{IJ}$ as
\begin{equation}
	f_{IJ}=\sqrt{\frac{(N_I+2N)(N_J+2N)}{N_I N_J}} \frac{\Gamma_{IJ}(-N_I-2N)}{\Gamma_{IJ}(N_I)}, \label{4.15}
\end{equation}
where we have shown explicitly the level-dependence in $\lambda^{-1}=\Gamma$. This corresponds to a generalization of the dualities obtained in \cite{Santos2023}. In particular, for the case of a single-WZW theory, this function recovers that one \cite{Kutasov1989}.

With this choice and identifying $U^I$ and $U^{I\dagger}$ with $g^I$ and $g^{I\dagger}$, respectively, we can write the dual partition function \eqref{4.12} in the form
\begin{eqnarray}
	Z_{SU(N)}&= & Z_g^{2}\int \mathcal{D} g^I\mathcal{D} g^{I\dagger}\exp \left[\sum_I\left(N_I+2N\right)\left(W[g^{I\dagger}]+W[g^I]\right)\right. \label{parti_func} \\ 
	&+&\left.4\pi\sum_{I,J}\sqrt{\left(N_I+2N\right)\left(N_J+2N\right)}\Gamma^{IJ}\left(-N_I-2N\right)\int \dd[2]{z} \tr J^I_L(g^\dagger)J^J_R(g)\right]. \nonumber 
\end{eqnarray}

So far, through non-Abelian bosonization, we have been able to decouple the full partition function \eqref{4.8} in subgroup sectors. The main piece that concerns us here is $Z_{SU(N)}$, which describes the competing interactions alluded to be responsible for the realization of the non-Abelian chiral spin liquid phases with boundaries housing gapless modes with central charge associated with the coset \eqref{0.2}. The starting point for this analysis can be either the model \eqref{bos1} or its dual \eqref{parti_func}.

%%%%%%%%%%%%%%%%%%%%%%%%%%%%%%%%%%%%%%%%%%%%%%%%%%%%%%%%%%%%%%%%%%%%%%%%%%%%%%%%%%%%%%%%%%%%%%%%%%%%%%%%%%%%%%%%%%%%%%%%%%%%%%%%%%%%%%%%%%%%%%%%%%%%%%%%%%%%%%%%%%%%%%%%%%%%%%%%%%%%%%%%%%%%%%

\section{Fixed Points} \label{Fixed_Points}
Now we are in a position to discuss the fixed points structure of the theory. To do so, we assume that $ \lambda_{1}=\lambda_{2}=\lambda $, $ \lambda^{t}_{1}=\lambda^{t}_{2} $,  and $ \lambda^{u}_{1}=\lambda^{u}_{2} $ for simplicity. Additionally, we set the regularization parameter $ b=c=1 $ in the equations \eqref{gamma} and \eqref{4.10} so that our bosonization procedure faithfully reproduces the emergent gauge invariance of the action \eqref{action with gauge fields} as $ \tilde{\Gamma}\rightarrow 0 $. Here, we focus on the fixed points for the $ SU(N) $ partition function. The remaining $ SU(N_{I}) $ and $ U(1) $ sectors exhibit fixed points only in the deep IR limit, where each sector is gapped with $ c=0 $, and the free fermion limit \cite{Santos2023}.
\subsection{Free Fermion}
As a consistency check, we verify that the UV free fermion fixed point is recovered from the bosonized theory. To this end, we initially take the limit $ \lambda_d\rightarrow 0  $ of the $ SU(N) $ partition function \eqref{parti_func}. In this limit, the matrix $ \lim\limits_{\lambda_d\rightarrow 0} \Gamma_{I,J}=\left(\frac{1}{\lambda}+1\right)\delta_{I,J} $, so that the partition function
\begin{align}
	\lim_{\lambda_d\rightarrow0}Z_{SU(N)}=Z_{g}^{2}\int \mathcal{D}g_{I}^{\dagger}\mathcal{D} g_{I} \exp\sum_{I} k_{I}\Biggl\{ W[g_{I}^{\dagger}g_{I}]+\frac{4\pi}{\lambda}\int \dd[2]{z}\tr J_{L}(g_{I}^{\dagger})J_{R}(g_{I})\Biggr\},\label{int_part_func_1}
\end{align}
where we defined $ k_{I}\equiv 2N+N_{I} $. Then, taking the limit $\lambda\rightarrow 0$ yields
\begin{align}
\lim_{\lambda,\lambda_{d}\rightarrow 0}Z_{SU(N)}=Z_{g}^{2}\int \mathcal{D}g_{I}^{\dagger}\mathcal{D} g_{I}  \left[\prod_{I}\delta(J_{L}^{I}(g^{\dagger}))\delta(J_{R}^{I}(g))\right]\exp\sum_{I}k_{I} W[g_{I}^{\dagger}g_{I}].
\end{align}

In order to perform the field integration, we use the property of the functional Dirac delta
\begin{align}
\delta(J_L(g^{\dagger}_{I}))\delta(J_R(g_{I}))=\frac{\delta(g^{\dagger}_I-\mathbf{1})\delta(g_I-\mathbf{1})}{\det(\partial_{z}\partial_{\bar z})_{adj}}, \label{delta property}
\end{align}
to obtain 
\begin{align}
	\lim_{\lambda,\lambda_{d}\rightarrow 0}Z_{SU(N)}=1.
\end{align}
Therefore, the $ SU(N) $ sector does not contribute to the total central charge of the system. With a similar reasoning, we obtain the same result for the $ SU(N_I) $ partition functions. At last, taking such limit in the $ U(1) $ sector yields the free fermion partition function, with central charge 
\begin{align}
	c_{UV}=N(N_{1}+N_2)
\end{align}

The remaining fixed points are found in the low energy limit, where the perturbations for the non-competing sectors are known to be relevant \cite{Huang2016,Santos2023}. For this reason, we assume that their respective coupling constants are in the strong coupling regime, such that each sector contributes with
\begin{align}
	c_{SU(N_{I})}=-\frac{N(N_{I}^{2}-1)}{N+N_{I}}\qquad \qquad \text{and} \qquad \qquad c_{U(1)}=N\left(N_{1}+N_{2}\right)-2 \label{other_central_charges}
\end{align}
to the total central charge of the model. A more detailed discussion on the $ U(1) $ sector central charge can be found in the Appendix \ref{AA}.

Another pair of fixed points arises when there are no competing interactions. These are expected when one coupling constant ($\lambda$ or $\lambda_d$) is fine tuned to zero and the other is left to flow to strong coupling. These cases reduce to the non-Abelian Thirring model we studied in \cite{Santos2023}.

\subsection{Non-Competing Gapped Fixed Point}
One such fixed point is found in the limit $ \lambda\rightarrow \infty $ and $ \lambda_{d}\rightarrow 0 $. Its partition function can be easily obtained from \eqref{int_part_func_1}
\begin{align}
\lim \limits_{\lambda\rightarrow\infty }\lim \limits_{\lambda_{d}\rightarrow 0}Z_{SU(N)}=Z_{g}^{2}\int \mathcal{D}g_{I}^{\dagger}\mathcal{D} g_{I} \exp\sum_{I} k_{I} W[g_{I}^{\dagger}g_{I}]. \label{part_func_trivial}
\end{align}
Its associated central charge,
\begin{align}
c_{SU(N)}=-\frac{N_{1}\left(N^{2}-1\right)}{N+N_{1}}-\frac{N_{2}\left(N^{2}-1\right)}{N+N_{2}},
\end{align}
confirms that we fully gapped the $ SU(N) $ sector. Adding the contributions from the $ SU(N_{I}) $ and $ U(1) $ sectors, equations \eqref{other_central_charges} and \eqref{u1 central charge}, the total central charge
\begin{align}
	c=c_{U(1)}+c_{SU(N_{1})}+c_{SU(N_{2})}+c_{SU(N)}=0.
\end{align}

This is the expected result, as by turning off the diagonal interaction our whole model reduces to a pair of decoupled fermions perturbed by current-current interactions. This is exactly the case of \cite{Santos2023}, where it was found that the non-Abelian Thirring interactions subtract the central charge associated with its WZW term.

In the complete quantum wires construction, see Figure \ref{Picture}, this phase supports gapless chiral degrees of freedom at the edges, as one chirality is left free at each edge. The central charge for these degrees of freedom
\begin{align}
	c=\frac{N_1(2^{2}-1)}{N_{1}+N}+\frac{N_2(N^{2}-1)}{N_{2}+N}
\end{align}
indicates that that statistic may be abelian or not depending on the specific choice of $ N_{1} $ and $ N_{2} $.

\subsection{Gapless Fixed Point}
Another non-competing fixed point if found in the limit $\lambda\rightarrow 0$, $\lambda_{d}\rightarrow \infty$, for which the partition function \eqref{parti_func} reduces to 
\begin{align}
Z_{g}^{2}\int \mathcal{D}g_{I}^{\dagger}\mathcal{D} g_{I}\delta \left[J_{L}(g_{1}^{\dagger})-J_{L}(g_{2}^{\dagger})\right]\delta\left[J_{R}(g_{1})-J_{R}(g_{2})\right] \exp \sum_{I} k_{I}W[g_{I}^{\dagger}g_{I}].
\end{align}
In order to integrate over the Dirac delta functions we use its reparametrization property once more,
\begin{align}
	\delta\left[J_{L}(g_{1}^{\dagger})-J_{L}(g_{2}^{\dagger})\right]\delta\left[J_{R}(g_{1})-J_{R}(g_{2})\right]&=\frac{\delta(g_{1}^{\dagger}-g_{2}^{\dagger})\delta(g_{1}-g_{2})}{\det (D_{z}D_{\bar z})_{adj}}\\ \nonumber
	&=Z_{g}^{-1}e^{-2N W[g_{1}^{\dagger}g_{1}]}\delta(g_{1}^{\dagger}-g_{2}^{\dagger})\delta(g_{1}-g_{2}),
\end{align}
which yields the partition function
\begin{align}
Z_{SU(N)}=Z_{g}\int \mathcal{D}g^{\dagger}\mathcal{D} g \exp \left(2N+N_{1}+N_{2}\right)W[g^{\dagger}g]. \label{competing z}
\end{align}
From its associated central charge 
\begin{align}
	c_{SU(N)}^{coset}=-\frac{(N_{1}+N_{2})(N^{2}-1)}{N_{1}+N_{2}+N},
\end{align}
we see that the diagonal interaction does not fully gap the $ SU(N) $ sector. In fact, it leaves some gapless degrees of freedom,
\begin{align}
	c=\frac{N_{1}N_{2}\left(2N+N_1+N_2\right)\left(N^{2}-1\right)}{(N_{1}+N)(N_{2}+N)(N_{1}+N_{2}+N)},
\end{align}
preserving a smaller conformal invariance, associated with the algebra
\begin{align}
	\frac{su(N)_{N_{1}}\times su(N)_{N_{2}}}{su(N)_{N_{1}+N_{2}}}. \label{coset}
\end{align}

In the complete quantum wires construction, one chirality for each bundle on the edge of the manifold is subject to only the diagonal interaction, see Fig. \ref{Picture}. In this way, the partition function in this limit is representative of the anyonic excitations expected on the edge of the quantum wires model.

\subsection{Competing Gapped Fixed Point}
At last, by taking the $\lambda\rightarrow \infty$ and $ \lambda_{d}\rightarrow \infty $ limit in \eqref{parti_func}, we obtain the partition function for the competing gapped fixed point 
\begin{align}
	Z_{SU(N)}= Z_{g}^{2}\int \mathcal{D}g_{I}^{\dagger}\mathcal{D} g_{I} \exp\sum_{I} k_{I}W[g_{I}^{\dagger}g_{I}]. \label{part_func_competing}
\end{align}
Notice that this partition function coincides with the one from the non-competing gapped fixed point, equation \eqref{part_func_trivial}. Thus, this is indeed a gapped fixed point
\begin{align}
	c=0.
\end{align}
Indicating that all the excitations in the bulk of the spin liquid have been fully gapped, leaving any dynamical behavior to the edge of the manifold, where the conformal degrees of freedom are associated with the algebra \eqref{coset}.

This coincidence is a consequence of our periodic boundary condition that eliminated the bundle structure of the quantum wires model \cite{Huang2016}. This greatly simplified our approach, allowing us to study the gap opening and RG flow in our model by considering the typical interactions in a bundle of the bulk. On the other hand, this prevents us from differentiating between the partition functions of the two gapped fixed points.

We highlight that, as we approach any of the IR strong coupling fixed points the current-current couplings that break conformal invariance vanish, so that the partition function is written in terms of only the products $ g_{I}^{\dagger}g_{I} $. In this limit, a new gauge invariance emerges, given by 
\begin{align}
	g_{I}^{\dagger} \rightarrow  g^{\dagger}_{I}\Lambda_{I}^{-1}(z,\bar z)\qquad \qquad &\text{and}\qquad\qquad g_{I}\rightarrow \Lambda_{I}(z,\bar z) g_{I}. \label{emergent_gauge_invariance}
\end{align}
At this point, is convenient to perform a variable change that makes this emergent gauge invariant explicit, we set $ g_{I}^{\dagger}g_{I}\rightarrow G_{I} $, such that the partition function contains an integration measure over an independent field, which is divergent.

These emergent gauge invariances take place in a very abrupt way, so that
there is a sudden change in the number of degrees of freedom that become unphysical due to the gauge redundancy. As we shall discuss, this leads to a discontinuity in the RG flow that manifests in the $ C $- and the $ \beta $-functions.

At last, we summarize the fixed point and central charge structure of the theory  in the table
\begin{align}	
	\begin{array}{c|cc}
		\text{central charge} &\lambda_1=\lambda_2=0& \lambda_1,\lambda_2\rightarrow\infty  \\
		\hline
		\lambda_{d}=0&N\left(N_{1}+N_{2}\right) & 0\\
		\lambda_{d}\rightarrow\infty& su(N)_{N_{1}}\times su(N)_{N_{2}}/su(N)_{N_{1}+N_{2}}&0
	\end{array} \label{fixed_points_table}
\end{align}

\section{The RG flow and the $ \beta $-functions} \label{RG flow}
With the fixed point structure at hand, we study the RG flow of our model in order to determine its phase diagram. To do so, we determine the $ \beta $-functions for the model and solve them to find the RG flow from a series of starting points in parameter space. The RG flow of the non-competing $ SU(N_{1}) $ and $ SU(N_{2}) $ can be found in the literature \cite{Santos2023}. There is no fixed point between the UV and the IR, such that starting at any point in parameter space the RG flows to strong coupling. As each sector is independent from the others, we only consider the RG flow generated by the $ SU(N)$ partition function \eqref{parti_func}.

Our bosonized partition function \eqref{parti_func}, is written in terms of a series of WZW actions perturbed by gap opening current-current interactions. The $ \beta $-function for this class of models has been extensively studied in the large level limit \cite{Georgiou2017, Georgiou:2018vbb, Delduc:2020vxy, Sfetsos:2013wia, georgiou2017integrable, Georgiou2017a, Georgiou2020, Georgiou2015}. 

In this section, we  follow the discussions and conventions found in the Appendix B of \cite{Delduc:2020vxy}. To do so, we recast our bosonized action \eqref{parti_func} into the form
\begin{align}
	S=-\sum\limits_{i,j}\frac{E_{i,j}}{8\pi}\int \dd[2]{x} \tr j^{i}_{\bar z}j^{j}_{z}+\sum_i \mathcal{K}_{i}S_{WZ}[h_{i}],
\end{align}
such that $ S_{WZ} $ is the last term of the WZW action \eqref{wzw action}, the ``currents''\footnote{Note that these coincide with the WZW $ SU(N) $ conserved currents for only one chirality for each $ g^{i} $.}  are given by 
\begin{align}
	j_{\mu}^{i}\equiv h^{-1}_{i}\partial_{\mu}h_{i}.
\end{align}
We set $ h_{i}\equiv\left(g_{1}^{\dagger},g_{1}^{-1},g_{2}^{\dagger},g_{2}^{-1}\right) $, which yields 
\begin{align}
		\mathcal{K}_{i}=\left( k_{1},- k_{1}, k_{2},- k_{2}\right),
\end{align}
and 
\begin{align}
	E_{i,j}=-\left(\begin{array}{cccc}
		 k_{1}& -2 k _{1}\Gamma_{1,1} & 0 &-2 \sqrt{ k_{1} k_{2}}\Gamma_{1,2}  \\
		0&  k_{1}  & 0 & 0 \\
		0 & - 2\sqrt{k_{1}k_{2}} \Gamma_{2,1} &  k_{2} & -2 k_{2}\Gamma_{2,2} \\
		0 & 0 & 0 &  k_{2}
	\end{array}\right).
\end{align}

Following the methods outlined in \cite{Delduc:2020vxy}, we find the first term in the large-$ \tilde k $ expansion of the $ \beta $-functions 
\begin{align}
	\beta_{\lambda_{1}}\equiv \dv{\lambda_{1}}{\ln \mu^{2}}=\frac{N X_{1}}{ k D}, \quad	\beta_{\lambda_{2}}\equiv \dv{\lambda_{2}}{\ln \mu^{2}}=\frac{N X_{2}}{ k_{2}D}, \quad \text{and} \quad \beta_{l_d}\equiv \dv{\lambda_{d}}{\ln \mu^{2}}=\frac{N X_{d}}{(k_{1}+k_{2}) D}, \label{beta func}
\end{align}
such that
\begin{align}
X_{1}&\equiv-2 k_1  k_2 \lambda _1 \left(\lambda _1+1\right) \left(2 \lambda _2+1\right)  \left(2 \lambda _d+2 \lambda _2+1\right) \left(\lambda _d+\lambda _1 \left(2 \lambda _d+\lambda _1+1\right)\right)\\ \nonumber
&- k_2^2 \lambda _1^2 \left(\lambda _1+1\right){}^2 \left(2 \lambda _d+2 \lambda _2+1\right){}^2- k_1^2 \left[\lambda _1^2\left(2 \lambda _2+1\right){}^2  \left(3 \lambda _d \left(\lambda _d+2\right)+1\right)\right.\\ \nonumber
&+\lambda _1 \lambda _d\left.\left(2 \lambda _2+1\right){}^2  \left(3 \lambda _d+2\right)+\lambda _2 \left(\lambda _2+1\right) \lambda _d^2+\lambda _1^4 \left(2 \lambda _2+1\right){}^2+2\lambda _1^3 \left(2 \lambda _2+1\right){}^2  \left(2 \lambda _d+1\right)\right],\\
X_{d}&\equiv-\lambda _d^2 \left[2 \left(2 \lambda _1+1\right) \left(2 \lambda _2+1\right) \left( k_1+ k_2\right) \lambda _d \left( k_2 \left(2 \lambda _1+1\right)+ k_1 \left(2 \lambda _2+1\right)\right)\right.\\ \nonumber
&+\lambda _d^2 \left( k_2 \left(2 \lambda _1+1\right)+ k_1 \left(2 \lambda _2+1\right)\right){}^2+\nonumber\\
&\left.\left(3 \lambda _1^2 \left(2 \lambda _2+1\right){}^2+3 \lambda _1 \left(2 \lambda _2+1\right){}^2+3 \lambda _2 \left(\lambda _2+1\right)+1\right) \left( k_1+ k_2\right){}^2\right],\\
D&\equiv\left[ k_1 \left(2 \lambda _2+1\right) \left(2 \lambda _d+2 \lambda _1+1\right)+k_2 \left(2 \lambda _1+1\right) \left(2 \lambda _d+2 \lambda _2+1\right)\right]^2. \label{beta_denominator}
\end{align}
Furthermore, $ X_{2} $ is given by replacing $  k_{1} \leftrightarrow  k_{2} $ and $ \lambda_{1}\leftrightarrow \lambda_{2} $ in $ X_{1} $, as expected.

Repeating this process for the other side of the duality, using the partition function \eqref{bos1}, we find that the $ \beta $-functions are described by the equations (\ref{beta func} - \ref{beta_denominator}) under the replacement $ k_{I}\rightarrow N_{I} $. 

This difference is due to the methods of \cite{Delduc:2020vxy}, which calculate the $ \beta $-functions in a large level expansion, resulting in a series in powers of $ N/N_{I} $ for the partition function \eqref{bos1} and $N/k_{I}$ for \eqref{parti_func}. As $k_{I}\equiv 2N+N_{I}$, we must take $ N_{I}>>N $ and the two $ \beta $-functions coincide in this limit.

In the weak coupling limit, the $ \beta $-functions read
\begin{align}
 \beta_{\lambda_{1}}&\approx-\frac{\lambda_{1}^{2}N}{k_{1}}-\frac{2\lambda_{1}\lambda_d N}{k_{1}+k_{2}}+\cdots, \label{beta_small}\\
 \beta_{d}&\approx-\frac{\lambda_{d}^{2}N}{k_{1}+k_{2}}+\cdots.
\end{align}
Which is the expected result from perturbative expansion. Notice that these $ \beta $-functions differ from those of \cite{Huang2016} by the last term in \eqref{beta_small}. This is due to the periodic contour conditions we imposed on our model that eliminated the bundle structure.

The beta functions we presented in equations \eqref{beta func} are divergent in the strong coupling limit. For the non-Abelian Thirring model it was found that the $ C $-function is discontinuous at the strong coupling fixed point, where there is a suddenly emergent gauge invariance \cite{Santos2023}. This is expected as the $ \beta $-function is proportional to derivatives of the  Zamolodchikov $ C $-function near the fixed points \cite{Zomolodchikov1986}. This discontinuity of the $ C $-function is a direct consequence of the decrease of degrees of freedom as the gauge invariance \eqref{emergent_gauge_invariance} suddenly emerges.

\begin{figure}
	\centering
	\includegraphics[width=0.7\linewidth]{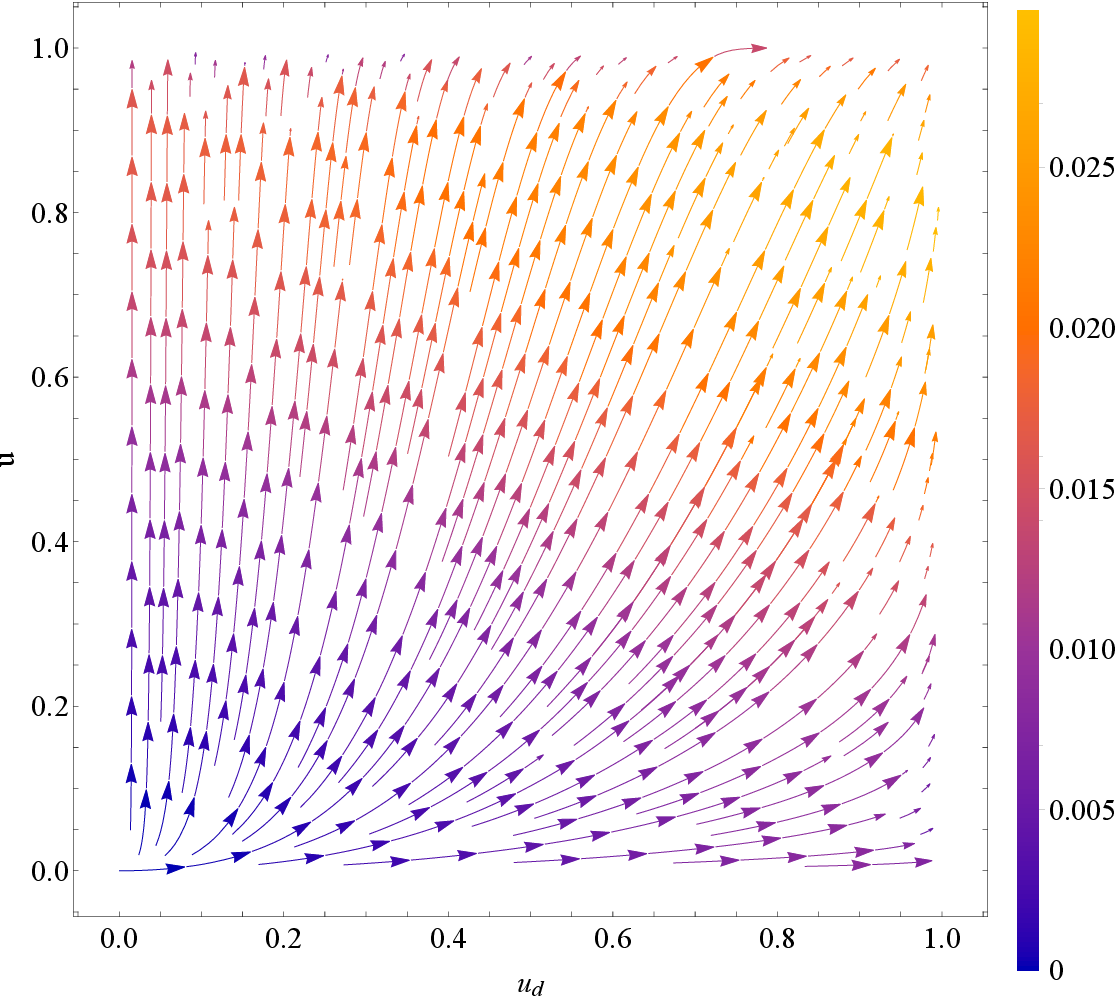}
	\caption{RG flow for $ N_{1}=N_{2}=20 $ and $ N=2 $.}
	\label{fig:fluxdiagramfull}
\end{figure}

\begin{figure}[h]
	\centering
	\includegraphics[width=0.7\linewidth]{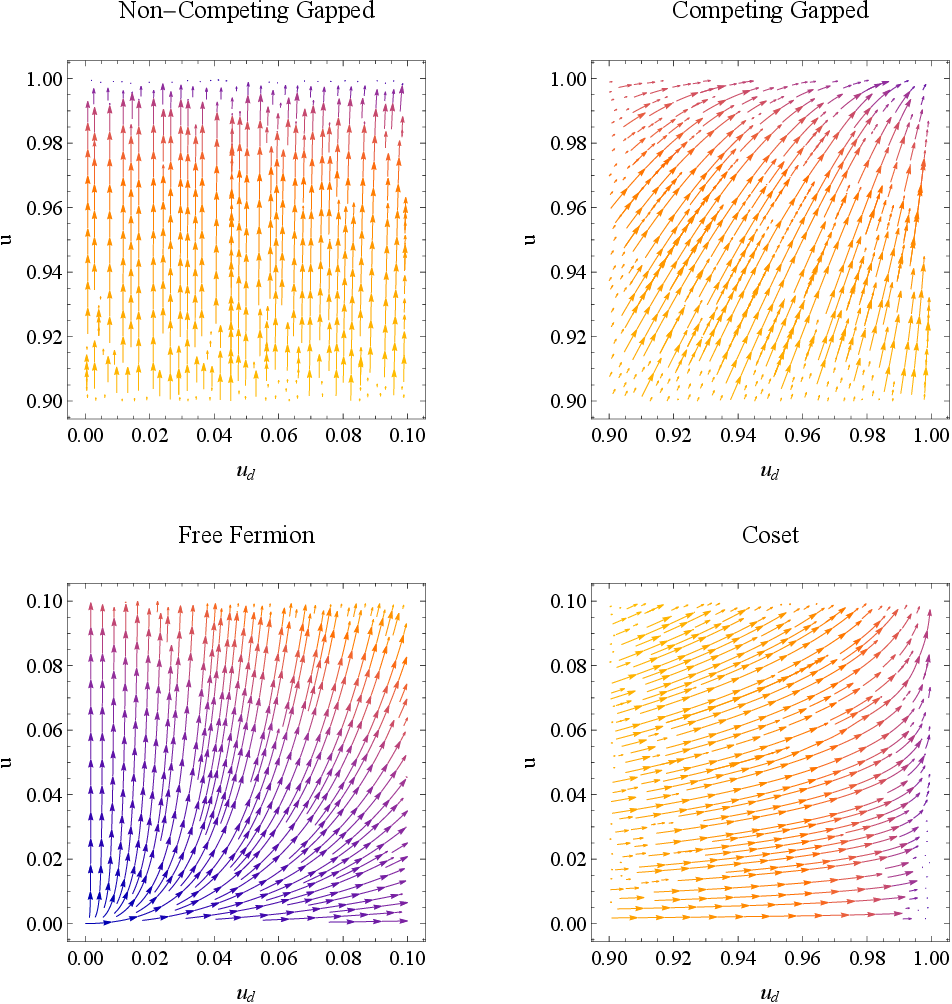}
	\caption{RG flow around the fixed points discussed in Section \ref{Fixed_Points}.}
	\label{fig:fixedpointsmosaic}
\end{figure}

In order to investigate the phase structure of the model, we need to study the RG flow in its entirety. To do so, it is convenient to compactify the space of coupling constants by introducing the reparametrizations
\begin{align}
	u_{1}=1-e^{-\lambda_{1}},\qquad u_{2}=1-e^{-\lambda_{2}}, \qquad \text{and}\qquad u_{d}=1-e^{-\lambda_{d}}. \label{u parametrization}
\end{align}
Under this new parametrization, the free fermion fixed point is given by $ u_{1}= u_{2} = u_{d} =0 $ and the strong coupling is achieved by flowing the new parameters to one, correspondingly. Furthermore, in order to keep our discussions concise, we will always present the results for $ \lambda_{1}=\lambda_{2} $. Nonetheless, they are independent coupling constants, and thus, they must be considered independently in calculations involving the RG flow, and then be taken to be equal when presenting our data. In terms of these new variables the full RG flux diagram is represented in Figure \ref{fig:fluxdiagramfull} and as a zoom into the fixed points in Figure \ref{fig:fixedpointsmosaic}.

It is important to call attention to the fact that our parametrization \eqref{u parametrization} compactifies the infinite length of the  RG flow for $ \lambda $ into a finite region. In this way, it obscures the notion of length in the parameter space, such that we cannot fully see the disconnection between the IR fixed points and the remaining RG flow that is found in the non-Abelian Thirring model \cite{Santos2023}. As a consequence of this disconnection, it is not possible to get to these fixed points exactly. But, we can get asymptotically close to them in a region of the parameter space where the physics is essentially indistinguishable from the fixed point.

Let us compare the diagram in Figure \ref{fig:fluxdiagramfull} with the fixed point structure found in Section \ref{Fixed_Points} by zooming in to the region around the fixed points, see Figure \ref{fig:fixedpointsmosaic}. The competing gapped fixed point is achievable by the majority of the parameter space as, when written in terms of the new variables, the $ \beta $-functions are only zero when their respective coupling constants are zero,
\begin{align}
	\beta_{\lambda}(u=0)=0 \qquad \text{and} \qquad \beta_{d}(u_d=0)=0,
\end{align}
and at their fixed points. In this way, the coset and non-competing gapped fixed points are only achievable by a fine tuning $ u=0 $ and $ u_{d}=0 $ respectively. 

\begin{figure}[h!]
	\centering
	\includegraphics[width=0.8\linewidth]{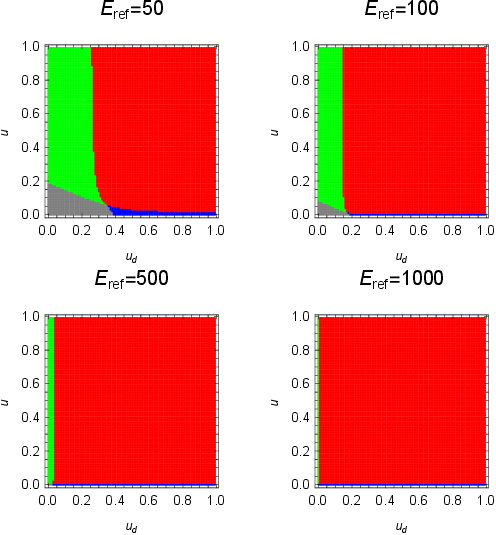}
	\caption{RG flow diagram for $ N=2 $, $ N_{1}=N_{2}=20 $, $E_{ref}$=50, 100, 500 and 1000 respectively.}
	\label{fig:phasediagram}
\end{figure}

This can be seen in the Figure \ref{fig:phasediagram}, where we plot a series of points as starting conditions for solving the $ \beta $-functions and color them according to their end points of the RG flow. Elaborating further, using the $ \beta $-functions, we find $ \lambda_1(\mu^{2}) $, $ \lambda_2(\mu^{2}) $ and $ \lambda_{d}(\mu^{2}) $ for each point in the diagram as a starting condition in the UV. Then, we take the IR limit of each solution and color their starting point in the diagram according to the rule: red if all the coupling constants flowed to strong coupling in the IR; green if only $ u $ flowed to strong coupling; blue if only $ u_{d} $ flowed to strong coupling; and grey if neither flowed to strong coupling. At last, we repeated this process for increasing values of $ E_{ref}\equiv\ln \mu^{2}_{UV}=-\ln\mu^{2}_{IR} $.

Notice that there is a thin line of blue points on the horizontal axis, corresponding to RG flows starting at $ u=0 $ that flow to the coset fixed point. Furthermore, the green region decreases as we increase the energy range, indicating that the non-competing gapped fixed point is only realized in the deep IR for a infinitesimally small region around the vertical axis. In this way, small perturbations from $ u=0 $ or $ u_{d}=0 $ destabilize these phases into the competing phase.

At last, the competing fixed point is realized in the deep IR for almost all the parameter space.

\section{Final Remarks} \label{final remarks}
Along this work we discussed the realization of topological phases of non-Abelian spin liquids.  To do so, we discussed how a quantum wires approach can effectively realize a (2+1)-dimensional phase in the deep infrared. The resulting model consists of a pair of fermions coupled by non-Abelian current-current Thirring-like interactions. One of the most direct ways to study the fixed point structure of this model is through a bosonization procedure, as the fixed points appear explicitly in the bosonic theory. However, this procedure is not unique. We discussed two different bosonization procedures, which yield two different, but of course equivalent, partition functions. This leads to an interesting result in itself. The bosonic version of the proposed model admits a remarkable strong-weak duality, mapping between the two partition functions.

We proceed by studying the RG flow of this model in a large level expansion. We found that the competing fixed point is achieved in the deep IR for almost every point in the parameter space, with the exception of a infinitesimally small region around $ \lambda_{d}=0 $ which decreases in size as we go deeper into the IR, and for $\lambda=0$. Thus, the QSL phase is realized in the IR strong coupling limit for most of the parameter space and is robust against perturbations. Conversely, the non-competing and coset phases are highly unstable as they depend on a fine tuning of the parameters to be realized in the deep IR. These results are best encapsulated by the diagram in Figure \ref{fig:phasediagram}.

Furthermore, the RG flow restricted to $\lambda=0$ can be thought of as representative of the sectors on the boundary bundles, see Fig. \ref{Picture}. Alternatively, this can be viewed as a model where there is no bulk bundle, leaving only the typical interactions of the boundary. By setting $\lambda=0$ we obtain a non-competing current perturbed WZW model similar to the one studied in \cite{Santos2023}. The conclusion is that $\lambda_{d}$ flows from the free fermion limit to strong coupling, where it realizes a new fixed point. At this point, a new conformal invariance described by the partition function \eqref{competing z} emerges, corresponding to gapless degrees of freedom on the boundary of the QSL phase.

We would like to underscore that these fixed points are not exactly realized for any finite energy scale. This is a consequence of the sudden change of degrees of freedom which is reflected in the $ C $ and $ \beta $-functions. This concept is made explicit by the Zamolodchikov metric for the non-Abelian Thirring model, where an asymptotic behavior is found near the strong coupling fixed points \cite{Kutasov1989,Georgiou2015}. This indicates that the RG flow acquires an asymptotic behavior near the fixed points, making them not reachable by an RG flow. Nonetheless, it is still possible to get arbitrarily close to the fixed points, in a configuration where the physics should be essentially indistinguishable.

%%%%%%%%%%%%%%%%%%%%%%%%%%%%%%%%%%%%%%%%%%%%%%%%%%%%%%%%%%%%%%%%%%%%%%%%%%%%%%%%%%%%%%%%%%%%%%%%%
\appendix

\section{Umklapp and Sine-Gordon Equivalence} \label{AA}

In this appendix we discuss the bosonization process of the Umklapp interaction. Since, the action in the partition function \eqref{4.9} consists of two decoupled similar parts, let us focus in just one of the pieces, for which the action can be taken as
\begin{align}
	S_{U(1)}&=\int d^2x\left[  \psi^{\dagger}_{L i \sigma}\left(\partial_{z}-iA_{z}\right) \psi_{L i \sigma} + \psi^{\dagger}_{R i \sigma}\left(\partial_{\bar{z}}-iA_{\bar{z}}\right) \psi_{R j \sigma}+\frac{1}{\lambda_t}A_{z}A_{\bar{z}}+\mathcal{L}_{Umklapp}\right], \label{2.11}
\end{align}
with $\mathcal{L}_{Umklapp}$ being of the form of one of the pieces (we take $I=1$) of \eqref{Umklapp_interactions}. We also set the total number of fermions $ \mathcal{N}=NN_{1} $

This form can be promptly bosonized after a suitable field redefinitions that partially decouples the vector fields from the fermions:
\begin{align}
	\psi_L\rightarrow e^{\theta+i\phi}{\psi}_L~~~&\text{and}~~~\psi^\dagger_L\rightarrow e^{-\theta-i\phi}{\psi}^\dagger_L\label{a2.13}, \\
	 \psi_R\rightarrow e^{-\theta+i\phi}{\psi}_R~~~&\text{and}~~~\psi^\dagger_R\rightarrow e^{\theta-i\phi}{\psi}^\dagger_R \label{a2.12},\\ 
	A_{z}=-i\partial_z\left(\theta+i\phi\right)~~~&\text{and}~~~A_{\bar{z}}=i\partial_{\bar{z}}\left(\theta-i\phi\right)\label{a2.14}. 
\end{align} 	
We recognize the change of variables (\ref{a2.13}) and (\ref{a2.12})  from the discussion of the $U(1)$ chiral anomaly. It leads to a nontrivial Jacobian of the form \cite{Fujikawa1980,Fujikawa1979} 
\begin{align}
	J _{F}^{U(1)}=e^{-\frac{\mathcal{N}}{\pi}\int d^2x\partial_z\theta\partial_{\bar{z}}\theta} \label{2.15}.
\end{align}
After the change of variables, we can put the partition function into the form
\begin{align}
	Z_{Umklapp}&=\int \mathcal D\mu\det\left(-\partial^{2}\right)\exp-\int \dd[2]{x}\left[\psi^{\dagger}_{L,i\sigma}\partial_{z}\psi_{L,i\sigma} +\psi^{\dagger}_{R,i\sigma}\partial_{\bar{z}}\psi_{R,i\sigma}+\frac{1}{\rho^{2}}\partial_{z}\theta\partial_{\bar{z}}\theta\right. \nonumber\\
	&\left.+\frac{1}{\sigma^{2}}\partial_{z}\phi\partial_{\bar{z}}\phi	 -\lambda_{u} \left(e^{-2\mathcal{N} \theta}\prod{\psi}_{R,i\sigma}^{\dagger} \prod{\psi}_{L,i\sigma} + e^{2\mathcal{N}\theta} \prod{\psi}_{L,i\sigma}^{\dagger} \prod{\psi}_{R,i\sigma}\right)\right], \label{2.16}
\end{align}
where the measure $\mathcal D\mu$ stands for $\mathcal D\theta \mathcal D\phi \mathcal D\psi_{R}\mathcal D \psi_L $,
\begin{equation}
	\rho^{2}\equiv \frac{\pi \lambda_{t} }{\pi+\lambda_{t}(\mathcal{N}+a)} \qquad \text{and}\qquad \sigma^{2}=\frac{\lambda_{t}}{1+a \lambda_{t}/\pi}.
\end{equation}

The term $\det\left(-\partial^{2}\right)$ in (\ref{2.16}) is due to the Jacobian for the transformations (\ref{a2.14}). Likewise, the  parameter $a$, which accompanies a local counterterm $B_zB_{\bar{z}}$, accounts for the possible ambiguities in the regularization of the fermionic Jacobian. 

%Now, let us count the number of $ U(1) $ degrees of freedom in this new action. The Jacobian cancels out the contribution from the bosons $\phi$ and $\theta$. This can be explicitly shown by writing the Jacobian in terms of ghost fields. For this cancellation to make sense in the $\lambda_t\rightarrow 0 $ limit, we impose that the kinetic terms contribution yields $\frac{1}{\det(-\partial^{2})}\delta (\theta) \delta(\phi)$ in the partition function. This constrains the vector fields to $ B_{z}=B_{\bar z}=0 $, making the kinetic terms zero, in a similar fashion to the free fermion limit of the $ SU(N) $ partition function \eqref{delta property}. Furthermore, subsequently taking the limit $ \lambda_{u}\rightarrow 0 $ yields the free fermion action, as expected.

At this point we are halfway through the $ U(1) $ bosonization process. As our last step we need to obtain the bosonized version of the Umklapp interaction, given by the last term of the partition function \eqref{2.16}. Our strategy closely follows the original discussion of the equivalence between the massless Thirring and the sine-Gordon models. In other words, we show that a series expansion of the partition function
\begin{align}
	Z=\int \mathcal{D} \psi\mathcal{D}\theta \exp -
& \int \mathrm{d}^2 x \left[{\psi}_{R, i\sigma}^{\dagger} \partial_z {\psi}_{R, i \sigma}+{\psi}_{L, i \sigma}^{\dagger} \partial_{\bar{z}} {\psi}_{L, i \sigma}+\frac{1}{\rho^2} \partial_z \theta \partial_{\bar{z}} \theta \right. \label{a2}\\
& -\lambda_u\left.\left(e^{-2 \mathcal{N} \theta} \prod {\psi}_{R, i \sigma}^{\dagger} \prod {\psi}_{L, i\sigma}+e^{2 \mathcal{N} \theta} \prod {\psi}_{L, i \sigma}^{\dagger} \prod {\psi}_{R, i \sigma}\right)\right]  \nonumber
\end{align}
in $\lambda_{u}$ is equivalent to an expansion of a sine-Gordon like partition function, given by
\begin{align}
	Z_{SG}=\int \mathcal{D}\vartheta \mathcal{D}\chi \exp -\!\int \! \dd[2]{x}\left[\chi_{R, i \sigma}^{\dagger} \partial_z \chi_{R, i \sigma} +\chi_{L, i \sigma}^{\dagger} \partial_{\bar{z}} \chi_{L, i \sigma}+ \frac{1}{\rho^{2}} \partial_{z}\vartheta \partial_{\bar z} \vartheta-\frac{\alpha_{0}}{\beta^{2}}\cos\left(\beta \mathcal{N}\vartheta\right)\right], \label{a1}
\end{align}
in powers of $\frac{\alpha_{0}}{\beta^{2}}$, provided there is a map between the coupling constants.

Before we do that, let us recall some well-known results for bosonic and fermionic free fields in 2D. Due to infrared divergences, it is convenient to add a regulator mass $m$ for $ \theta $, which should be taken to zero in the end of our calculations. In this way, the Feynman propagator for the massive field $\theta$ is then given in terms of the modified Bessel function
\begin{align}
	\Delta_F\left(m;\mu;x-y\right)=\frac{\rho^2}{2\pi}K_0\left(m\left|x-y\right|\right)-\frac{\rho^2}{2\pi}\ln\left(\frac{\mu}{m}\right),\label{2.31}
\end{align}
where  $\mu$ is an arbitrary mass parameter that we inserted in order to keep the Feynman propagator regular in the $ m\rightarrow 0 $ limit
\begin{align}
	\Delta_F\left(0;\mu;x-y\right)&=\lim_{m\rightarrow 0}\left[\frac{\rho^2}{2\pi}K_0\left(m\left|x-y\right|\right)-\frac{\rho^2}{2\pi}\ln\left(\frac{\mu}{m}\right)\right]\nonumber\\
	&=\lim_{m\rightarrow 0}\left[-\frac{\rho^2}{2\pi}\ln\left(cm\left|x-y\right|\right)-\frac{\rho^2}{2\pi}\ln\left(\frac{\mu}{m}\right)+\mathcal O (m^{2})\right]\nonumber\\
	&=-\frac{\rho^2}{2\pi}\ln\left(c\mu\left|x-y\right|\right),\label{2.32}
\end{align}
where $c$ is a numerical constant (related to the Euler constant).
The UV divergences are regularized by a large mass cutoff $\Lambda$ in such way that $\Delta^{reg}_F\left(m;\mu;\Lambda;x\right)\equiv\Delta_F\left(m;\mu;x\right)-\Delta_F\left(\Lambda;x\right)$ has a finite $x\rightarrow 0$ limit
\begin{align}
	\Delta^{reg}_F\left(m;\mu;\Lambda;0\right)=-\frac{\rho^2}{2\pi}\ln\left(\frac{\mu}{\Lambda}\right).\label{2.33}
\end{align}

On the other hand, the fermions do not need any regularization and thus we have
\begin{align}
	S_F\left(x-y\right)\equiv\langle\chi_L(x)\chi^\dagger_L(y)\rangle_0=-\frac{i}{2\pi}\frac{1}{z-w}\label{2.34}
\end{align}
and
\begin{align}
	\bar{S}_F\left(x-y\right)\equiv\langle\chi_R(x)\chi^\dagger_R(y)\rangle_0=-\frac{i}{2\pi}\frac{1}{\bar{z}-\bar{w}},\label{2.35}
\end{align}
with $z=x_0+ix_1, \bar{z}=x_0-ix_1$ and $w=y_0+iy_1, \bar{w}=y_0-iy_1$.

At this point, it is convenient to consider the correlation function of a general vertex operator, this can be easily derived from the regularized propagator
\begin{align}
	\langle e^{i\sum \beta_i\theta(x_i)}\rangle_0&=\exp\left[-\frac{1}{2}\sum_{i,j}\beta_i\beta_j\left(\Delta_F\left(m;\mu;x_i-x_j\right)-\Delta_F\left(\Lambda;x_i-x_j\right)\right)\right]\nonumber\\
	&=\left(\frac{m}{\mu}\right)^{\left(\rho^2/4\pi\right)\left(\sum\beta_i\right)^2}\left(\frac{\mu}{\Lambda}\right)^{\left(\rho^2/4\pi\right)\sum\beta^2_i}\prod_{i<j}\left(c\mu\left|x_i-x_j\right|\right)^{\left(\rho^2/2\pi\right)\beta_i\beta_j},\label{2.36}
\end{align}
where the subscript is to mean that the expected value should be taken with respect to the partition function \eqref{a2} with $ \lambda_{u}=0 $. Furthermore, we call attention to the fact that this correlation function vanishes unless $ \sum\limits_{i}\beta_{i}=0 $. Thus, we can expand the Umklapp partition function \eqref{a2} as
\begin{align}
	Z_{Umklapp}&=\sum^{\infty}_{k=1}\frac{\left(\lambda_u\right)^{2k}}{\left(k!\right)^2}\left[\int\prod^{k}_{j=1}\dd[2]{x_j}\dd[2]{y_j}\right]\langle \exp{-2\mathcal{N}\sum^{k}_{j=1}\left[\theta(x_j)-\theta(y_j)\right]}\rangle_{0~bosonic}\nonumber\\
&\times\langle\prod^{k}_{j=1}\left[\prod_{i,\sigma}{\psi}^\dagger_{R,i,\sigma}\left(x_j\right)\prod_{i,\sigma}{\psi}_{L,i,\sigma}\left(x_j\right)\prod_{i,\sigma}{\psi}^\dagger_{L,i,\sigma}\left(y_j\right)\prod_{i,\sigma}{\psi}_{R,i,\sigma}\left(y_j\right)\right]\rangle_{0~fermionic}.\label{2.37}
\end{align}

Furthermore, we rearrange the fermion fields, such that, we rewrite the correlation functions appearing in this partition function as
\begin{align}
	&\prod_{i,\sigma}\langle\prod^{k}_{j=1}\left[{\psi}_{R,i,\sigma}\left(y_j\right){\psi}^\dagger_{R,i,\sigma}\left(x_j\right)\right]\prod^{k}_{j=1}\left[{\psi}_{L,i,\sigma}\left(x_j\right){\psi}^\dagger_{L,i,\sigma}\left(y_j\right)\right]\rangle_0\nonumber\\
&=\langle\prod^{k}_{j=1}\left[{\psi}_{R}\left(y_j\right){\psi}^\dagger_{R}\left(x_j\right)\right]\rangle^{NN_{1}}_0\langle\prod^{k}_{j=1}\left[{\psi}_{L}\left(x_j\right){\psi}^\dagger_{L}\left(y_j\right)\right]\rangle^{\mathcal{N}}_0,\label{2.38}
\end{align}
where we used that the fermions are identical, thus we use anyone of the $ \psi_{R/L,i,\sigma} $ (which we simply call $ \psi_{R/L} $) to express the whole correlation function. At this point, we use the Wick theorem and the fermion propagators (\ref{2.34}) and (\ref{2.35}) to obtain the correlation function
\begin{align}
	\langle\prod^{k}_{j=1}\left[{\psi}_{R}\left(y_j\right){\psi}^\dagger_{R}\left(x_j\right)\right]\rangle_0&=\left(\frac{-i}{2\pi}\right)^k\sum_{p\in S_k}\text{sign}\left(p\right)\prod^k_{j=1}\frac{1}{\bar{w}_i-\bar{z}_{pj}}\nonumber\\
	&=\left(-1\right)^k\det\left(\frac{1}{\bar{z}_i-\bar{w}_{k}}\right)\nonumber\\
	&=\left(\frac{-i}{2\pi}\right)^k\left(-1\right)^k\left(-1\right)^{k\left(k-1\right)/2}\frac{\prod_{1\leq i<j\leq k}\left(\bar{z}_i-\bar{z}_{j}\right)\left(\bar{w}_i-\bar{w}_{j}\right)}{\prod^k_{i,j=1}\left(\bar{z}_i-\bar{w}_{j}\right)}.\label{2.39}
\end{align}
Analogously,
\begin{align}
	\langle\prod^{k}_{j=1}\left[{\psi}_{L}\left(x_j\right){\psi}^\dagger_{L}\left(y_j\right)\right]\rangle_0&=\left(\frac{-i}{2\pi}\right)^k\left(-1\right)^{k\left(k-1\right)/2}\frac{\prod_{1\leq i<j\leq k}\left(z_i-z_{j}\right)\left(w_i-w_{j}\right)}{\prod^k_{i,j=1}\left(z_i-w_{j}\right)}.\label{2.40}
\end{align}
Finally, we combine the results (\ref{2.36}) and (\ref{2.38})-(\ref{2.40}) into (\ref{2.37}) and obtain
\begin{align}
	Z_{Umklapp}&=\sum^{\infty}_{k=1}\frac{\left(\lambda_u^{(r)}\left(\mu c\right)^{\mathcal{N}}\right)^{2k}}{\left(k!\right)^2}\left[\int\prod^{k}_{j=1}\dd[2]{x_j}\dd[2]{y_j}\right]\left(\frac{1}{2\pi}\right)^{2k}\nonumber\\
	&\times\frac{\prod_{1\leq i<j\leq k}\left(c^2\mu^2\left|x_i-x_{j}\right|\left|y_i-y_{j}\right|\right)^{2\mathcal{N} \left(1-\rho^2\mathcal{N}/\pi\right)}}{\prod^k_{i,j=1}\left(c\mu\left|x_i-y_{j}\right|\right)^{2\mathcal{N}\left(1-\rho^2\mathcal{N}/\pi\right)}},\label{2.41}
\end{align}
where we introduced the renormalized coupling constant
\begin{align}
	\lambda_u^{(r)}=\lambda_u\left(\frac{\mu}{\Lambda}\right)^{-\rho^2\mathcal{N}/\pi}.\label{2.42}
\end{align}

On the other hand, by expanding the SG partition function \eqref{a1} in powers of $\alpha_0$, we get
\begin{align}
	Z_{SG}=\sum^{\infty}_{k=1}\frac{1}{\left(k!\right)^2}\left(\frac{\alpha_0}{\beta^2}\right)^{2k}\left[\int\prod^{k}_{j=1}\dd[2]{x_j}\dd[2]{y_j}\right]\langle \exp{i\mathcal{N}\beta\sum^{k}_{j=1}\left[\vartheta(x_j)-\vartheta(y_j)\right]}\rangle_{0},\label{2.43}
\end{align}
where once again, we used the neutrality condition for the correlation functions of vertex operators. Using the general result (\ref{2.36}), we obtain
\begin{align}
	Z_{SG}=\sum^{\infty}_{k=1}\frac{1}{\left(k!\right)^2}\left(\frac{\alpha}{\beta^2}\right)^{2k}\left[\int\prod^{k}_{j=1}\dd[2]{x_j}\dd[2]{y_j}\right]\left(\frac{1}{2\pi}\right)^{2k}\frac{\prod_{1\leq i<j\leq k}\left(c^2M^2\left|x_i-x_{j}\right|\left|y_i-y_{j}\right|\right)^{\mathcal{N}^{2}\beta^2\rho^{2}/2\pi}}{\prod^k_{i,j=1}\left(cM\left|x_i-y_{j}\right|\right)^{\mathcal{N}^{2}\beta^2\rho^{2}/2\pi}},\label{2.44}
\end{align}
with
\begin{align}
	\alpha=\frac{\alpha_0}{2}\left(\frac{M}{\Lambda}\right)^{\beta^2\mathcal{N}^{2}\rho^{2}/4\pi},\label{2.45}
\end{align}
and $M$ is an arbitrary mass parameter which is the equivalent of $\mu$ plays the same role for $\vartheta$ that $\mu$ plays for $\theta$.

Comparing (\ref{2.41}) with (\ref{2.44}), we conclude that $Z_{Umklapp}=Z_{SG}$ provided we make the identifications
\begin{align}
	\mu&=M,\qquad\qquad\quad\qquad~~ \frac{\beta^2}{4\pi} =\frac{1+a\lambda_{t}/\pi}{\mathcal{N}\lambda_{t}},\nonumber\\
	\frac{\alpha}{\beta^2}&=\lambda_u^{(r)}\left(c\mu\right)^{\mathcal{N}}\qquad 
	\text{and}\qquad \rho^{2}\equiv  \frac{\pi \lambda_{t}}{\pi+\lambda_{t}(\mathcal{N}+a)}. \label{2.46}
\end{align}
This establishes the equivalence between the $ U(1) $ partition function \eqref{4.9} and the partition function
\begin{align}
	Z_{U(1)}=Z_{g}^{2}\int \mathcal{D} \vartheta_{I}\mathcal{D} \chi_{I} \exp- \int \dd[2]{x}\sum_{I}&\left[\chi_{R,i \sigma}^{I,\dagger}\partial_{z}\chi_{R,i \sigma}^{I}+\chi_{L,i \sigma}^{I,\dagger}\partial_{\bar z}\chi_{L,i \sigma}^{I}\right. \\ \nonumber &\left.+\frac{1}{\sigma^{2}}\partial_{z}\phi_{I}\partial_{\bar z}\phi_{I}+\frac{1}{\rho^{2}}\partial_{z}\vartheta_{I}\partial_{\bar z} \vartheta_{I} -\frac{\alpha_{0}}{\beta^{2}}\cos(\beta N \vartheta_{I})
	\right],
\end{align}
where $ Z_{g}=\det-\partial^{2} $ is the $ U(1) $ ghost term.

The free fermion limit of the $ U(1) $ partition function is achieved by taking $ \alpha_{0}/\beta^{2},\sigma $ and $ \rho $ to zero, while $ \beta\rightarrow\infty $. This produces a factor $ \delta(\vartheta)\delta(\phi)/\det(-\partial^{2}) $, such that when integrating over the boson fields we are left with the free fermion partition function,  in a similar fashion to the free fermion limit of the $ SU(N) $ partition function \eqref{delta property}. On  the other hand, the strong coupling limit is achieved by taking $\alpha_{0}/\beta^{2}\rightarrow \infty$, while $ \beta, \rho $ and $ \sigma $ assume finite regularization dependent values. This process gaps the $\vartheta$ bosons and leaves $ \phi $ gapless \cite{mussardo}. Then, adding the contributions from the fermions and the ghosts, the central charge for the strong  coupling fixed point reads
\begin{align}
	c_{U(1)}=N(N_{1}+N_{2})-2. \label{u1 central charge}
\end{align}

\bibliographystyle{ieeetr}
\bibliography{spin_liquids}	

\end{document}